# Hubble's law and Superluminity Recession Velocities

Author: Leonid S. Sitnikov
E-mail: Lsitnikov@cs.com

## Abstract

The extension of the so-called "empty" (with gravity and antigravity that compensate each other in full or do not exist at all) universe and cosmological redshift in it are considered in this paper. Its flat space-time can be submitted not only as manifold with Friedman-Robertson-Walker metrics (FRW) of the general theory of relativity (GR) but also as space-time with usual Minkowski metrics (M-metrics) of the special theory of relativity (SR); the transfer of metrics can be done by suitable transformation of reference frame. Both below-mentioned statements are equally fair for such the universe. First: the distant galaxies can have superluminity recession velocities in FRW—space of GR; we have no right to use here the formula of relativistic Doppler effect. Secondly: the SR theory is fair in the M-space and, accordingly, recession velocities of the same galaxies here can aspire to the speed of light only.

In this article it is shown that, despite opposite pictures in FRW-and M-spaces, in the careful account of all details both approaches yield results agreed among them. Thus, actually there are no contradictions between the interpretations of cosmological redshift, based on SR and GR.

The comment: 7 pages, 3 figures, 7 bibliographic references, Copyright 2005.

## Two approaches to redshift

Hubble's law was discovered as a result of observations. It has an approximate character like each empiric law. It establishes the existence of linear dependence of recession velocities of far galaxies and distances to them. We cannot measure directly parameters that appear in Hubble's law: velocities and distances. In reality we determine supernova brightness, which gives us information about distance, and redshift $Z=\Delta\lambda/\lambda$ in spectrum of its radiation. Hubble has found out the correlation between brightness and parameter $Z$. In the time of discovery and development of Hubble's law it was acceptable to explain redshift phenomenon by relativistic effect of Doppler in the context of SR, i.e. it was accepted to associate redshift $Z$ with recession velocity of a galaxy. In this case, however, Hubble's dependency is believed to be linear, roughly only, and with $Z<1$. When $Z$ and distance are increasing unrestricted, the recession velocity achieves asymptotically the light speed. SR doesn't allow observing galaxies that recede with $V=>C$ because in this case $Z=>\infty$.

However, in modern theoretical cosmology it is conventional to interpret the redshift in the frames of GR and general cosmological principle (the Universe is homogeneous, isotropic and looks identical at a given time to any observer). It causes Friedman's cosmological models, i.e. space-time with FRW-metrics. The analysis of the expansion of the Universe is done in resent articles [1-3]. Strictly speaking, such analysis was done at the dawn of becoming the relativistic cosmology and given in any encyclopedia or book that touches problems of cosmology (see for example, [4,5]. However, the important feature of articles [1-3] is, that their results of the analysis lead up to the logic end – to the precisely formulated and strictly proved statement that in FRW-space the distant galaxies scatter with velocity, which can be equal to the speed of light or exceed this speed. After reading papers [1-3] and then reviewing to earlier publications, it becomes easy to see that each of them contains actually all preconditions, necessary to come to the same conclusion (so, in [5] within the framework of GR it is deduced that there is strictly linear dependence of recession velocity of galaxies from distance, which isn't limited).

Let us write down FRW-metrics in the simplified form:

$$ds^2 = -c^2 dt^2 + R^2(t) d\chi^2$$

Here $\chi$ - commoving distance and $R(t)$ - the scale factor; it is accepted usually that its size is equal to the radius of the Universe. Then dimensionless scale factor $R(t_2/t_1)$ is attitude $R(t_2)/R(t_1)$ of radiuses of the Universe in two of its various states. It shows, that all proper distances between objects with zero peculiar speeds have grown in $R(t_2/t_1)$ time from the moment of $t_1$ up to the moment $t_2$. Expression "all distances" includes also lengths of waves of light; therefore the size of redshift Z in GR is completely defined by the dimensionless scale factor, namely

$$Z+1 = R(t_2)/R(t_1)$$

The product of the scale factor on the commoving coordinate $\chi$ is proper distance $D=R(t)\chi$ in our usual understanding [1-3]. The derivative of this distance with respect to time is velocity of recession of the distant galaxies:

$$V = dD/dt = R'(t)\chi = R'(t)D/R(t) = H(t)D$$

Thus, strong linear dependence between recession velocity and distance ($H(t)=R'(t)/R(t)$ is Hubble's constant or, more precisely, Hubble's parameter) is deduced directly from GR.

Whereas, mentioned expressions derived theoretically on the base of GR, there is good reason to believe that they have the most fundamental nature. It means that Hubble's law metamorphosed from experimental and rough law to a fundamental and perfect one. A conclusion follows from this new law of such a high level [3-5] that, contrary to SR, the recession velocity can achieve speed of light on a large enough distance, or, with still longer distance, it can go beyond this limit (speed of light). At the same time the value of $Z$ remains quite finite. World lines of receding galaxies are described by dashed lines in figures 1,2. The horizontal stroke-dotted straight-line $t=t_H$ represents spacelike section of space-time at the moment "now", and the curve - a light cone of the past.

### How to coordinate GR and SR approaches

We will try to solve the stated problem, comparing positions of [1-3] with results of the analysis of the process in the space-time with the usual Minkowski metrics (M-metrics), which can be formed from FRW-metrics by suitable transforming of the reference frame.

For this purpose we will proceed from the above-mentioned general expression for $ds^2$ to the metrics of the open isotropic model

$$ds^2 = c^2 t^2 - R^2(t)[d\chi^2 + sh^2\chi(d\theta^2 + \sin^2\theta d\varphi^2)]$$

Also we will consider its special case with scale factor $R(t)=ct$ (it corresponds to expansion of the universe that is free from matter [5]). Today we also can generalize this expression regarding the "empty" universe, in which gravity and antigravity counterbalance each other.

So, when $R(t)=ct$, we will find

$$ds^2 = c^2 t^2 [1 - d\chi^2 + sh^2\chi(d\theta^2 + \sin^2\theta d\varphi^2)]$$

In space with such metrics the recession velocity of galaxies $V=dD/dt=R'(t)\chi=c\chi$ grows beyond all bounds in process of increase of $\chi$. In this respect such model is the same as general FRW-models with superluminity velocities (despite of its space is flat).

And now let us change the last expression for $ds^2$ by formulas $r=ct*sh\chi$, $\tau=t*ch\chi$ (see [5] and then [6,7]). As a result we will have

$$ds^2 = -c^2 d\tau^2 + dr^2$$

– the standard metrics of Minkowski space.

Let us compare now the features of the searching space with $R(t)=ct$ in the coordinates $t$-$\chi$ or $t$-$D$ (Fig.1,2) and in the coordinates $\tau$-$r$ (Fig.3). First of all, it is easy to see from the comparison of figures, that each hyperbolic subspace of the Fig.3 is in keeping with some spacelike section ($t=t_i$) of the Fig.2. It means that increasing of flowing time $t$ on the Fig.1,2 corresponds with the following: the points of any _hyperbole_ in the Fig.3 pass to another hyperbole with bigger value $s$ (but not only from one horizontal straight line to another). In other words, the process of expanding of the Universe is reflected in the reference frame $\tau$-$r$, as scanning of our usual Minkowski space not by a horizontal straight line of flowing time $t$ ($t=t_1$, $t=t_2$,..), as we count up to now, but by a hyperbole of running interval $s$ ($s=s_1$, $s=s_2$,..).

Then, we remember, that the relation of the scale factors defines value of redshift Z in space with FRW-metrics:

$$Z+1 = R(t_2)/R(t_1) = ct_2/ct_1 = t_2/t_1, \quad \text{i.e.} \quad Z+1 = t_2/t_1$$

What will take place with value Z in derivative M-space? As $\tau=t*ch\chi$, for it

$$Z+1 = t_2/t_1 = (\tau_2/ch\chi)/(\tau_1/ch\chi) = \tau_2/\tau_1$$

Thus, value Z does not change at transition from one space to another.

It is different with the recession velocity. While we have in FRW-space (for the empty Universe in coordinates $t$-$D$)

$$V = dD/dt = c\chi,$$

in M-space

$$U = dr/d\tau = d(ct*sh\chi)/d(t*ch\chi) = c*th\chi$$

So, in FRW-space the recession velocity can be smaller, equal or greater the speed of light $c$, depending on size of accompanying coordinate $\chi$, but in derivative M-space the recession velocity of the same galaxies always remains smaller than speed of light, aspiring to $c$ in process of increase $\chi$ infinitely.

And, at last, let us discuss situation with the light cone. For an extending photon $ds=0$, therefore $cdt=R(t)d\chi$ (it follows directly from FRW-metrics). For our case when scale factor $R(t)=ct$, it means that

$$\chi(t)=\ln(t_H/t)=\ln(Z+1),$$

or, that is the same,

$$t=t_H \exp(-\chi)=t_H/(ch\chi+sh\chi)$$

The light cone of the past, appropriate to the first expression, is submitted on Fig.1,2. For its translation in M-space we use equations $r=ct*sh\chi$ and $\tau=t*ch\chi$. And as a result we will receive the formula of isotropic straight line (Fig.3) as it must be in M-space

$$\tau=\tau_H-r/c$$

On the other hand, coming back to the previous expression, we can copy it as

$$t_H/t=ch\chi+sh\chi=ch\chi(1+th\chi)=(1+U/c)/(1-U^2/c^2)^{1/2}=Z+1$$

In this way the equation of the light cone in FRW-space completely repeats the formula of the relativistic Doppler's effect. As a result of such recurrence we can work as in M-space with Doppler's effect and with the isotropic straight line, so in FRW-space without Doppler's effect but with the light cone of the complicated form. In any case we get the same value of redshift $Z$.

Now it is obvious, that, having measured $Z$ of a distant galaxy, we can calculate all interesting parameters by two various ways. First, we can measure value $Z$ and count coordinates $t,\chi$ and the superluminal (for enough big $Z$) velocity of a galaxy in FRW-space

$$\chi(Z)=\ln(Z+1), \qquad t=t_H \exp(-\chi), \qquad V=dD/dt=c\chi,$$

And after that with the help of the expressions $r=ct*sh\chi$, $\tau=t*ch\chi$ we are able to turn to M-space and to define coordinates $r,\tau$ and the velocity $U=c*th\chi$ of the same galaxy in this space. It is obvious that the velocity $V$ becomes equal to the speed of light, if $\chi=1$, i.e. $Z=\exp(1)-1=1.72$. At the same time the velocity $U$ equals $0.76c$. When $Z=10$, we will get $V=2.4c$ and $U=0.98c$, and when $Z=1000$, the velocities will be $V=6.9c$ and $U=(1-2*10^{-6})c$.

Secondly, it is possible to act on the contrary and to find at once speed of the galaxy in space with M-metrics

$$U=[(Z+1)^2-1]/[(Z+1)^2+1]$$

The point of crossing of the hyperbole $c^2\tau^2-r^2=c^2\tau_H^2$ with the world line $r=U\tau$ of the galaxy will give coordinates of the last one. And now we will find easily position and speed of the galaxy in FRW-space.

## Discussion of results

So, we have considered the "empty" universe (specific kind of universe). According to GR, such the universe should expand with the constant speed that is equal to the speed of light, i.e. $R(t)=ct$. There realizes in our Universe or there does not this case - it is completely unimportant in the context of the given work. Another factor is important: the "empty" universe does not differ from the universe 30/70, in sense of existence of superluminity recession velocities. Such velocities should exist in FRW-space for all universes; and SR, including relativistic Doppler's effect, turns out inapplicable.

We know also, that a completely other physical scenario turns out by consideration of the same process (of the expansion of the "empty" universe) as recession of galaxies (trial balloons) in the reference frame $\tau$-$r$, i.e. in the M-metrics: the biggest recession velocities do not achieve the speed of light; their values are defined unequivocally with the help of the formula of relativistic Doppler's effect.

And at last, we were convinced that both pictures not only do not contradict to each other, but on the contrary, each picture follows from another. Certainly, such situation should cause feeling of discomfort and, moreover, some bewilderment. But only until we will have recollected the closest connection of physics with geometry of space-time: it is impossible to speak about physics in a separation from geometry. The considered here problem of interpretation of cosmological redshift is a bright illustration of importance of this well-known statement. Now it becomes clear, that all disputes and misunderstanding around of the discussed problem are related to following: one of the arguing sides holds in mind, as self-evident and consequently not taken out for discussion, FRW-geometry of the space-time in the coordinates $t$-$D$, while the second side does the same but with geometry of the usual M-space, i.e. in the coordinates $\tau$—$r$. Thus each of them is completely right within the framework of its geometry. But to come to the united point of view and to establish full co-ordination of received results - is impossible for them. There is not a real trifle for this purpose: it is necessary to realize, formulate in an obvious kind and to inform the opponent the representations about geometry of space – time, which your physical model is created and developed in.

In conclusion, let us concern two more problems. First, can we apply the transfer of metrics and reference frame, which was used by us for the "empty" expanding universe, to other models of the Universe? It is clear that the isometric transfer, which we used, does not change the internal curvature of the FRW-space. So we don't get Minkowski space in the general case of FRW-space. However, let us formulated this problem differently: can we transfer from the reference frame $t$-$D$ to the reference frame $\tau$—$r$ in general case of FRW-space? It becomes obvious with such a formulation: yes, we can. The recessing galaxies, which lie on the spacelike surface $t=t_i$, in the coordinates $t$-$D$, will lie in this case not on the hyperboloid in M-space but on more complicated surface in the coordinates $\tau$—$r$. The last surface will be different for each model of the Universe. We can do such a transfer by the simplest graphical method – dot by dot from Fig.2 to Fig.3. The recession velocity of any far galaxy will not exceed the speed of light in reference frame $\tau$—$r$. Thus, the model of the "empty" universe is not distinguished in regarding the velocities.

Second: which combination "geometry+physics" (from the two combinations that we have considered) is closer to reality and therefore is preferable for us? We were already convinced, that measurements of value Z do not allow us to make a choice; both combinations are equally competent and lead to the results agreed among them. Certainly, it does not exclude that any of systems can have the certain advantages in the researching of some concrete problems, be (or appear) more habitual or useful. So, it is shown in each of the papers [6,7] that GR and SR descriptions of the empty universe are equivalent, however, one of them is preferable (SR in [6],GR – in [7]). However, it seems that just the possibility to describe the "empty" universe in the different reference frames is the most interesting feature of such a universe. As a result, we have now the chance to compare the descriptions and to achieve greater comprehension of some cosmological problems. If this will be a reality, the "empty" universe model could take the place of the Rosetta stone in cosmology.

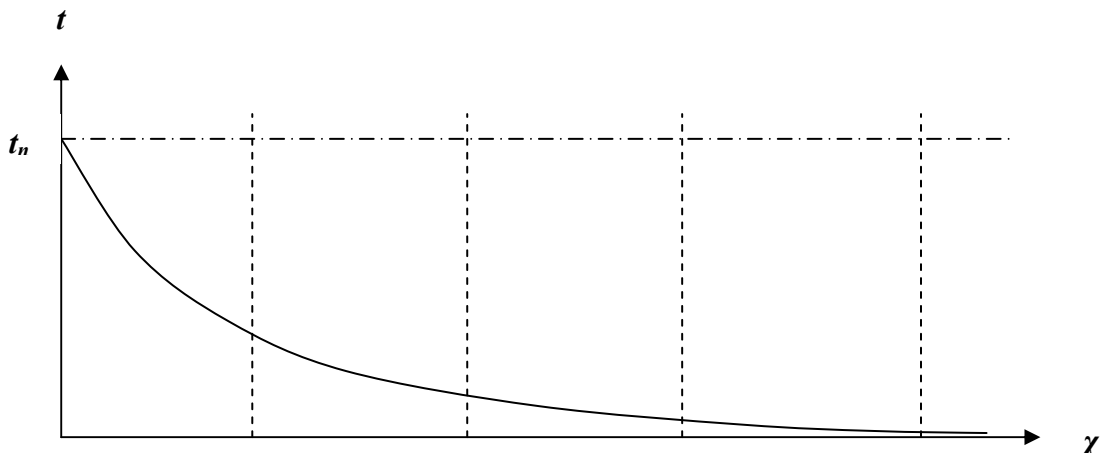

Fig.1

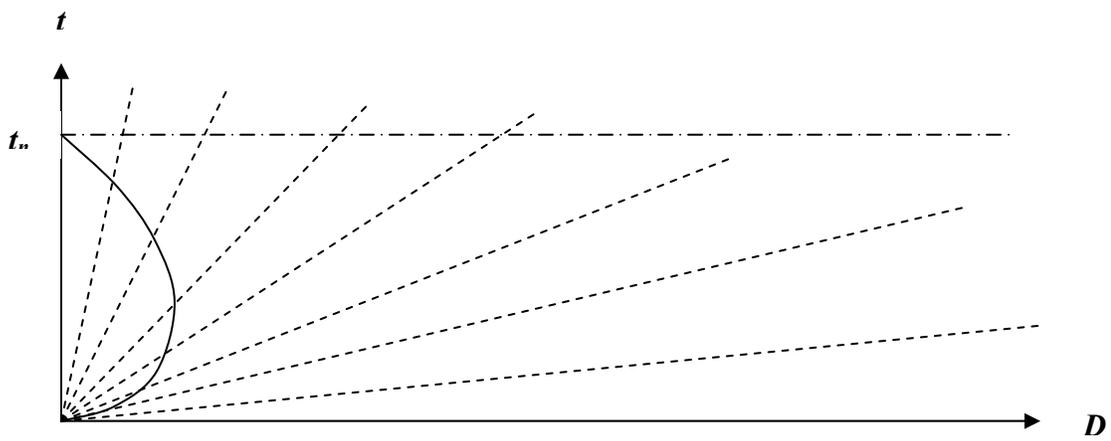

Fig.2

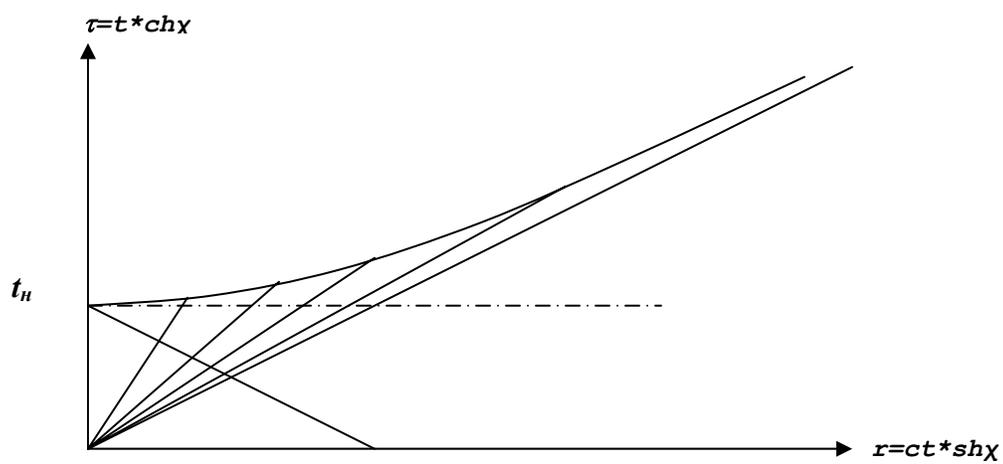

Fig.3